\documentclass[11pt]{article}
\pdfoutput=1

\usepackage{amsmath, amsfonts, amssymb}
\usepackage{comment}
\usepackage{graphicx}
\usepackage{psfrag}
\usepackage[usenames,dvipsnames,svgnames,table]{xcolor}
\usepackage{enumerate}
\usepackage{jheppub}
\usepackage{soul}
\usepackage{subfig}

\newcommand{\be}{\begin{equation}}
\newcommand{\ee}{\end{equation}}

\renewcommand{\d}{\textrm{d}}
\newcommand{\e}{\textrm{e}}

\newcommand{\SU}{\mathop{\rm SU}}

\newcommand{\ba}{\begin{eqnarray}}
\newcommand{\ea}{\end{eqnarray}}

\renewcommand{\d}{\textrm{d}}

\newcommand{\diag}{\textrm{diag}}

\def\Im           {{\rm Im\hskip0.1em}}

\def\calk         {{\cal K}}

\def\caln         {{\cal N}}

\def\calw         {{\cal W}}

\title{Compact G2 holonomy spaces from SU(3) structures}
\author[\clubsuit]{S. Andriolo,}
\author[\spadesuit]{G. Shiu,}
\author[\heartsuit]{H. Triendl,}
\author[\bigstar]{T. Van Riet,}
\author[\bigstar]{G. Venken,}
\author[\clubsuit,\diamondsuit]{and G. Zoccarato}

\emailAdd{sandriolo@connect.ust.hk, shiu@physics.wisc.edu, h.triendl@imperial.ac.uk, thomas.vanriet, Gerben.Venken @kuleuven.be, gzoc@sas.upenn.edu}

\affiliation[\clubsuit]{\it Center  for  Fundamental  Physics  and  Institute  for  Advanced  Study,
Hong  Kong  University  of  Science  and  Technology,  Hong  Kong}
\affiliation[\spadesuit]{\it Department of Physics, University of Wisconsin-Madison, Madison, WI 53706, USA}
\affiliation[\heartsuit]{\it Department of Physics, Imperial College London, London SW7 2AZ, UK}
\affiliation[\bigstar]{\it Instituut voor Theoretische Fysica, KULeuven, Celestijnenlaan 200D, B-3001 Leuven, Belgium  }
\affiliation[\diamondsuit]{\it Department of Physics and Astronomy, University of Pennsylvania, Philadelphia, PA 19104, USA}

\begin{document}

\abstract{We construct novel classes of compact G2 spaces from lifting type IIA flux backgrounds with O6 planes.  There exists an extension of IIA Calabi--Yau orientifolds for which some of the D6 branes (required to solve the RR tadpole) are dissolved in $F_2$ fluxes. The backreaction of these fluxes deforms the Calabi-Yau manifold into a specific class of SU(3)-structure manifolds. The lift to M-theory again defines compact G2 manifolds, which in case of toroidal orbifolds are a twisted generalisation of the Joyce construction. This observation also allows a clear identification of the moduli space of a warped compactification with fluxes. We provide a few explicit examples, of which some can be constructed from T-dualising known IIB orientifolds with fluxes. Finally we discuss supersymmetry breaking in this context and suggest that the purely geometric picture in M-theory could 
provide a simpler setting 
to address some of
the consistency issues of moduli stabilisation and de Sitter uplifting. }

\rightline{\small MAD-TH-18-07, Imperial/TP/18/HT/02}

\maketitle

\newpage
\section{Introduction}

Riemannian manifolds with special holonomy are of interest to both mathematicians and physicists. In physics they appear in string- and M-theory as compactification manifolds that preserve some fraction of the supersymmetry in 10 or 11 dimensions. The moduli spaces then correspond to the space of massless scalar fields in the effective lower-dimensional field theory and the moduli-space metric determines the kinetic term for these scalars. In order to stabilise (some of) these moduli one can switch on fluxes in the internal manifold. The 10-dimensional Einstein equations then imply a non-trivial curvature being induced. If one insists again on preserving supersymmetry then the Ricci-flat special holonomy condition changes into a special structure group condition. It is said that the manifold has a certain $G$-structure, where $G$ is the structure group. This has been nicely reviewed in \cite{Grana:2005jc, Koerber:2010bx}.

Aside from fluxes, further stabilisation of moduli can come from quantum corrections to the effective action. In this paper we ignore the latter and we investigate classical 11-dimensional backgrounds with a 7-dimensional manifold of compact G2 holonomy. One of our aims is to provide new constructions of such manifolds. Our method relies on using the duality between IIA string theory and M-theory. Since we work entirely at the classical level, we simply rely on the fact that 11d supergravity, reduced on a circle becomes 10d IIA supergravity and that there is a precise way to interpret certain singularities in 11d as stacks of branes in 10d.

Manifolds with compact G2 holonomy have been discussed before in several works, most notably the original work of Joyce \cite{Joyce1, Joyce2} and more recently many more examples were found by using twisted connected sums \cite{Kovalev:2001zr}. The reader can consult \cite{Grigorian:2009ge} for a fairly-recent review and \cite{Halverson:2014tya, Braun:2017uku, Kennon:2018eqg} for some most recent state-of-the art work on constructing these manifolds and their relation with various 10-dimensional string theories using dualities. The construction method presented in this paper is however different.

Before we discuss the actual details of our construction, we present the main idea underlying our construction. For that we first use the observation of Kachru and McGreevy \cite{Kachru:2001je} that  compact G2 spaces obtained from desingularising certain orbifolds of $\mathbb{T}^7$ have a simple IIA dual description in terms of IIA Calabi-Yau (CY) orientifolds.  
Compact G2 spaces also admit Type IIA Calabi-Yau orientifold descriptions where the Calabi-Yau manifold is smooth but the branes are intersecting \cite{Cvetic:2001tj,Cvetic:2001nr,Cvetic:2001kk}. In fact, these constructions \cite{Cvetic:2001tj,Cvetic:2001nr} are prototypes of the intersecting brane constructions of realistic particle physics models, whose M-theory lift to G2 spaces was discussed in \cite{Cvetic:2001kk}.
More generally, one can discuss Calabi-Yau 3-folds $X$ that allow discrete involution symmetries that leave special Lagrangian 3-cycles invariant such that they can be promoted to Calabi-Yau orientifold solutions of IIA string theory. The compactness of the CY orientifold enforces a RR tadpole condition which implies the addition of D6 branes wrapping cycles in the same homology class as the O6 planes such that the total 6-brane charges vanishes. Such manifolds should lift to compact 7-dimensional manifolds $Y$ with G2 holonomy  of the form
\begin{equation}\label{CY}
Y = \frac{X \times S^1}{\mathbb{Z}_2}\,,
\end{equation}
where the involution acts both on the M-theory circle and the Calabi-Yau $X$, and the singularities of Y are resolved after the ${\mathbb{Z}_2}$ orbifolding. Explicit realisations of this are discussed in \cite{Kachru:2001je}.

The main point of our construction is the observation that the IIA background $X$ can be deformed with fluxes while preserving supersymmetry and the flatness of the 4D vacuum (Minkowski), in such a way that the lift to 11 dimensions provides again a compact G2 space. For that to be possible the only allowed fluxes are the RR two-form flux $F_2$ since any other flux would switch on the M-theory 4-form destroying the flatness of the 7d compact space. More loosely speaking one could say that the D6 branes in the Calabi-Yau constructions get dissolved in $F_2$ fluxes, which then backreact the CY geometry $X$ into a specific (non-Ricci flat) $\SU(3)$-structure $\tilde{X}$ in such a way that the lift to 11d provides a Ricci flat compact G2 space $\tilde{Y}$. Hence the 7d space $\tilde{Y}$ can be seen as a twisted product (non-trivial fibre product) of a  $\SU(3)$-structure $\tilde{X}$ with a circle, where the twist is such that the 7d geometry $\tilde{Y}$ remains Ricci flat:
\begin{equation}\label{Twisted}
\tilde{Y} = \frac{\tilde{X} \ltimes S^1}{\mathbb{Z}_2}\,.
\end{equation}
To our knowledge the first examples of these ``generalised Calabi-Yau'' compactifications of IIA on $\tilde{X}$ were found in \cite{Kachru:2002sk} by T-dualising type IIB orientifolds with 3-form fluxes and a general discussion of such solutions appeared in \cite{Grana:2006kf}.
An important difference between \eqref{Twisted} and \eqref{CY} is that in \eqref{Twisted} the ${\mathbb{Z}_2}$ involution may by construction have less fixed points or even acts freely, depending on the number of D6-branes. Each singularity introduces degrees of freedom that can be associated with their resolution. The non-trivial information about RR two-form flux and (some of the) O6-planes is completely encoded in the non-trivial fibration of $S^1$ over $\tilde{X}$. This fibration is not unique and in one-to-one correspondence with the possible set of consistent, supersymmetry-preserving $F_2$ fluxes one can put over $\tilde{X}$. In principle this method works for any BPS combination of sources. For simplicity however we restrict to parallel D6/O6 sources.

In the course of this work we learned that 
this research direction is not entirely uncharted, as some initial forays in the case of 
non-compact G2 spaces can be found
in the references \cite{Kaste:2002xs, Kaste:2003dh, Gaillard:2009kz, Fayyazuddin:2006gr}. However, to our knowledge, we are the first to point out how this method allows the construction of new \emph{compact} spaces of G2 holonomy, provide an investigation of the resulting moduli spaces, and suggest how to break supersymmetry.

The phenomenologically interesting properties of our new examples is that the G2 spaces $\tilde{Y}$ have smaller moduli spaces than their counterparts $Y$, since in the IIA picture we obviously stabilise some moduli using the $F_2$ flux on $\tilde{X}$. However, concerning interesting matter-couplings, we have co-dimension 4 singularities that can accommodate non-Abelian gauge groups, but not chiral matter. This is because in this paper, we consider M-theory lift of IIA backgrounds with parallel O6/D6 sources.  It would be interesting to extend our analysis include backgrounds with both fluxes and non-parallel sources, like the ones in \cite{Marchesano:2004yq,Marchesano:2004xz} as in this case the corresponding G2 lifts would have desirable phenomenological features. However, we will not attempt to do so here because of technical complications with the supergravity solutions. What we find especially appealing in our constructions is that our G2 moduli spaces should be identical to the IIA(IIB) moduli spaces of warped compactifications with flux, after using the M-IIA duality (and IIB T-duality in cases it is known). The latter is notoriously complicated \cite{Giddings:2005ff, Douglas:2008jx, Shiu:2008ry} due to fluxes and the loss of a purely geometric picture. What we show here is that for a certain set of type II flux vacua a pure geometric picture in terms of G2 holonomy spaces exists.\footnote{See \cite{Schulz:2012wu} for similar lines of reasoning.} This might lead to a deeper understanding of warped effective field theories.

\section{SU(3) structures from IIA strings}

\subsection{General discussion}
We start from the general metric Ansatz in 10D string frame
\begin{equation}\label{10Dmetric}
\d s^2_{10} = \e^{2W} \eta_{\mu\nu}\d x^{\mu}\d x^{\nu} + \d s^2_6\,.
\end{equation}
Aside from the metric, the possible non-zero supergravity fields are the 2-form fieldstrength $F_2$ and the dilaton $\Phi$ that is allowed to depend on the 6D coordinates. Supersymmetry of flux backgrounds is best understood using the pure spinor formulation \cite{Grana:2005jc}, which in this context boils down to the following two poly-form equations
\begin{align}
& \d (e^{2W-\phi}\Phi_+)=0\,,\nonumber\\
& \d (e^{2W-\phi}\Phi_-) = e^{2W-\phi}dW\wedge\bar{\Phi}_- -\frac{i}{8}e^{3W}\star_6 F_2\,.\label{poly}
\end{align}
To obtain compactifications to 4d Minkowski space orientifolds are unavoidable. From here onwards we assume as only sources parallel O6 planes and possibly D6 branes. One can show that, in the presence of an O6 plane, the pure spinors $\Phi_{-}, \Phi_{+}$ define an $SU(3)$-structure as follows
\begin{equation}
\Phi_- = \frac{e^W}{8}\Omega\,, \qquad \Phi_+ = \frac{e^{i\theta}e^W}{8}e^{-iJ} \,.
\end{equation}
where  $\theta$ just denotes an irrelevant phase. The real 2-form $J$ and complex 3-form $\Omega$ are essential data of the $SU(3)$-structure. Our conventions for $\SU(3)$-structures are spelled out in Appendix \ref{App:conventions}.

The polyform equations (\ref{poly}) are solved when:
\begin{align}
\label{susy:phi(W)}
& e^{\phi - 3W} = e^{\phi_0} = \text{constant}\equiv g_s    \,,	\\
\label{susy:J}
& \d J = 0								\,,		\\
\label{susy:OmegaR}
& \d (e^{-W} \Omega_R) = 0			\,,		\\
\label{susy:OmegaI}
& \d (e^W \Omega_I) = - g_s e^{4W} \star_6 F_2	\,.
\end{align}
The last equation should be interpreted as the one that defines $F_2$ in terms of the $SU(3)$ structure and the warping. To promote this to a solution of the 10D EOM, it is sufficient to enforce the Bianchi identity \cite{Koerber:2007hd}
\begin{equation}
\d F_2 =  \sum_iQ_i \delta_i \,,
\end{equation}
where the right hand side symbolises all localised sources carrying 6-brane charges. Using the pure spinor equations the Bianchi identity amounts to
\begin{equation}
\d \Bigl(e^{-4W}\star_6 \d(e^W \Omega_I) \Bigr) = - g_s \sum_iQ_i \delta_i\,.
\end{equation}
The $F_2$ field strength is obtained from \eqref{susy:OmegaI} as
\begin{equation}
\label{F_2fromOmegaI}
g_s F_2 = - e^{-4W}\star_6 \d(e^W \Omega_I) \,.
\end{equation}
We can split $F_2$ into the sum of a ``background part'' and a ``sourced part'' $F_2 = F^B_2 + F_2^S$, where the background part is defined as the value of $F_2$ in the constant warp factor limit (or smeared limit), that is 
\be
F^B_2 \sim \star_6 \d\Omega_I^{smeared} \ .
\ee
The remaining part of $F_2$ is accounted by $F^S_2$.

The only non-zero torsion classes for these orientifold solutions with just $F_2$ flux are $W_2$ and $W_5$
\begin{equation}
\label{W2_F2}
W_2 = i e^\phi  F^{1,1}\,,\qquad \d W = W_5 + \bar W_5 \,.
\end{equation}
The computation of the torsion can be found in Appendix \ref{App:torsion}.

From \eqref{W2_F2} one can deduce that $F_2^B$ has to be $(1,1)$ and primitive. This is indeed the supersymmetry condition on the fluxes in case the orientifold is smeared and this can be explicitly verified using 4d supergravity.\footnote{Note that the standard equations for the 4d supergravity ignore the warping.} On the other hand $F_2^S$ can have in general $(1,1)$, $(2,0)$ and $(0,2)$ pieces, but the $(1,1)$ part has to be primitive.

\subsection{An example from its type IIB mirror}
An explicit IIA flux backgrounds of the above kind was found in  \cite{Grana:2006kf, Kachru:2002sk} by T-dualising three times a simple IIB orientifold with 3-form fluxes. The triple T-duality is such that the $H_3$ flux is gone and replaced by ```metric flux" whereas $F_3$ is T-dualised to $F_2$ and the O3/D3 sources are mapped to parallel O6/D6 sources.

Let us follow the example in \cite{Grana:2006kf} (see also \cite{Kachru:2002sk} ) and strip off all details, such as moduli dependence and flux/charge quantisation, and warping since we simply want to illustrate the essence. The internal space is $\mathbb{T}^6/\mathbb{Z}_2$ with $\mathbb{Z}_2$ the orientifold involution. The pure spinors are
\begin{align}
 e^{-i J} &=(1+i  e^6\wedge e^1)\wedge (1-i e^5 \wedge e^2) \wedge (1-2i e^4 \wedge e^3)\,,\\
\Omega &= (e^6-i e^1)\wedge (e^5 +i e^2 ) \wedge (e^4 + 2 i e^3) \equiv dz_1 \wedge dz_2 \wedge dz_3\,,
\end{align}
where $e^i$ are the Cartesian one-forms $d x^i$. The fluxes are given by
\begin{align}
g_s G_3 = g_s F_3 - i H_3 = \frac{i}{2} \left[d\bar z_1\wedge dz_2 \wedge dz_3 +dz_1 \wedge d\bar z_2 \wedge dz_3 -2 dz_1 \wedge dz_2 \wedge d\bar z_3\right]\,,
\end{align}
such that $G_3 \in H^{(2,1)} (\mathbb{T}^6)$. Therefore the solution preserves at least $\mathcal{N}=1$ supersymmetry. It is possible to show that this solution preserves not more supersymmetry,  but in absence of further orbifolding this vacuum can be described as an $\mathcal N=1$ state in an $\mathcal{N}=4$ gauged supergravity.

A triple T-duality along $x_4, x_5, x_6$ gives the nilmanifold defined by the following Maurer-Cartan equations:
\begin{align}
\d e^4 = e^{12}\,,\qquad \d e^5 = e^{13}\,,\qquad\d e^6 = e^{23}\,,
\end{align}
whereas $e^1, e^2$ and $e^3$ are closed (and simply given by $d x^i$). The flux and pure spinors become
\begin{align}
&\Omega =  (e^1 + ie^6) \wedge (e^2 - ie^5) \wedge (e^3 -2i e^4)
\label{ex:J}\,,\\
& J = e^1 \wedge e^6 - e^2 \wedge e^5  - 2 e^3 \wedge e^4 \,,\\
& g_s F_2 = - 4 e^{34} + e^{25} - e^{16}\,.		
\end{align}
This solution solves the equations of motion with smeared O6 sources. The solution with localised sources is derived in the Appendix \ref{App:localised}.

This example of T-dualising O3 backgrounds to O6 backgrounds with just $F_2$ flux can be generalised, and we give more examples in section \ref{sec:examples}. But the general picture is rather straightforward: the mirror of the D3/O3 sources leads to D6/O6 sources (possibly non-aligned), wrapping torsional cycles (see below). The mirror of H-flux leads to an $\SU(3)$-structure with torsion. Whereas the mirror of the $F_3$-flux is the background (non-sourced part of the) $F_2$ flux. The mirror of the $F_5$ flux is the sourced part of the $F_2$ flux. The mirror of the ISD condition is then the Hodge duality between the torsion $\d\Omega_I$ and $F_2$.

\subsection{Comments on RR tadpoles}
\label{sec:IItad}

Tadpole cancellation conditions are crucial for achieving compactness of the flux solutions. RR tadpoles are simply the statement that compact spaces allow no net charges. From a mathematical viewpoint these conditions ensure that the PDE's describing the G2 space (after lift to 11D, see next section), allow solutions describing compact spaces. In practice tadpole cancellation conditions may be understood as necessary conditions to have a well-defined solution to the Bianchi identities of the NSNS and RR fields. In the more familiar case of type IIB compactifications with O3/D3 sources the tadpole cancellation condition comes from the Bianchi identity of the RR five-form of type IIB which has the form\footnote{We define $\tilde F_p = d C_{p-1} - H \wedge C_{p-3} + m e^B$ where $m$ is the Romans mass.}
\begin{align}
\label{eq:BIF5} d \tilde F_5 = H_3 \wedge F_3 + 2 \kappa_{10}^2\, T_3 \, \rho_{\text{D3}} \,,
\end{align}
where $\rho_{\text{D3}}$ contains the contributions to the D3-brane charge coming from various sources in the compactification (that is O3/D3 and/or D7-branes with magnetic fluxes or wrapping curved 4-cycles).\footnote{We choose the convention of measuring the charge of the orientifold plane in the space quotientied by the orientifold involution, which leads to the relation $Q($Op$)=-2^{p-5} Q($Dp).} Given that the internal space has a non-trivial topology it is necessary to impose some conditions to ensure that a solution to \eqref{eq:BIF5} exists with a well defined profile for $\tilde F_5$. For the case at hand this is equivalent to asking that the right hand side of the Bianchi identity \eqref{eq:BIF5} is zero in cohomology, a condition which translates into the familiar equation
\begin{align}
\frac{1}{2 \kappa_{10}^2 T_3} \int H_3 \wedge F_3 + Q_{\text{D3}} = 0 \,,
\end{align}
where $Q_{\text{D3}}$ is the integral of the D3-brane charge density.

We can now follow this discussion and translate it to type IIA compactifications with O6/D6 sources, where sources couple magnetically to the RR two form $\tilde F_2$ so there is only one Bianchi identity that we need to check. Moreover it is important to note that in a twisted torus differential forms can be non-closed even when they have constant coefficients. Let us review that: using the Maurer-Cartan forms $e^i$ we can express any differential $p$-form $\Upsilon$ with constant coefficients as
\begin{align}
\Upsilon = \frac{1}{p!} \Upsilon_{i_1 \dots i_p}\, e^{i_1} \wedge \dots \wedge e^{i_p}\,.
\end{align}
However given that the Maurer-Cartan forms are not closed the differential form written above is not automatically closed even if the coefficients of the tensor are. In fact taking the exterior derivative we have\footnote{The covering space of a twisted torus is a group manifold and in our conventions the Maurer--Cartan forms on group manifolds obey $\d e^i = -\tfrac{1}{2}f^i_{jk}e^j\wedge e^k$.}
\begin{align}
\d \Upsilon = -\frac{1}{2(p-1)!} f_{jk}^l \Upsilon_{l i_1 \dots i_{p-1} } \,e^j \wedge e^k \wedge e^{i_1} \wedge \dots \wedge e^{i_{p-1}}\,.
\end{align}
In the following we employ a shorthand notation writing the right hand side of the previous equation as $f \cdot \Upsilon$, giving therefore that $\d\Upsilon = f \cdot \Upsilon$. Using this we can easily obtain the Bianchi identity for $\tilde F_2$
\begin{align}\label{eq:BIF2}
\d\tilde F_2^S = - f \cdot \tilde F^B_2 + 2 \kappa_{10}^2\, T_6\, \rho_{\text{D6}}\,,
\end{align}
where we have split $\tilde{F}_2$ in its background $\tilde F^B_2$ and sourced part $\tilde F_2^S$ as discussed below equation (\ref{F_2fromOmegaI}). To make a comparison with \eqref{eq:BIF5} we have that under T-duality $\tilde F_5$ maps to $\tilde F_2^S$, $F_3$ maps to $\tilde F_2^B$ and $\rho_{\text{D3}}$ maps to $\rho_{\text{D6}}$. Again we can pass this equation and convert it into a tadpole cancellation condition by asking that the right hand side is a trivial class in cohomology. However it is not  sufficient to ask that the right hand side vanishes in de Rham cohomology (that is in $H^3_{dR}(\tilde X) \simeq H^3(\tilde X,\mathbb R)$) but it is necessary to impose this condition in integer cohomology (that is in $H^3(\tilde X,\mathbb Z)$). The difference between the two is relevant for us in the following because in our examples all sources are parallel and wrap torsional cycles and therefore their contribution to the Bianchi identity would be invisible in de Rham cohomology.   The Poincar\'e dual $\Pi$ of this homology class will therefore be a torsion element of $H^3(\tilde X,\mathbb Z)$ and we take it to be of order $N$, that is $[N \Pi] =0$.\footnote{Here we used standard Poincar\'e duality, i.e. $H^p(\mathcal M,\mathbb Z) \simeq H_{d-p}(\mathcal M,\mathbb Z)$ for $\text{dim}(\mathcal M) =d$. In addition to this it is sometimes useful to recall that by using the universal coefficient theorem it is possible to prove that $\text{Tor}\, H_p(\mathcal M,\mathbb Z)\simeq \text{Tor}\, H^{p+1}(\mathcal M,\mathbb Z)$. For the precise statements and proofs one may look at Chapter 3 of  \cite{Hatcher:478079}.} Moreover to obtain the tadpole we need to consider what happens to the term $f \cdot \tilde F_2$. However given the way we constructed it, $f \cdot \Upsilon$ is always an exact form, and therefore it disappears when taking \eqref{eq:BIF2} in cohomology. Putting all of this together we find that the tadpole condition is \cite{Marchesano:2006ns}
\begin{align}\label{eq:D6t}
N_{\text{D6}} + r N = 2 N_{\text{O6}}\,, \qquad \exists\, r \in \mathbb Z\,.
\end{align}
The meaning of this equation is that the total charge carried by the sources, that is the D6-branes and the O6-planes need to be a multiple of $N$ to ensure that it is trivial in cohomology. This implies that the total charge D6-brane charge is a well defined object only modulo $N$, meaning that there can be different solutions satisfying the RR tadpole condition with different number of D6-branes. In particular it is possible to have transitions between different solutions via some domain-wall where $N$ D6-branes dissolve into flux \cite{Maldacena:2001xj,Cascales:2003zp},\footnote{Even if these domain-walls are present it may be impossible to nucleate them, for instance in supersymmetric cases, thus effectively making the number of D6-branes stable.} a phenomenon observed also in GKP solutions for the D3-brane charge. This seems to give a large freedom in building the solutions, specially given the fact that the number $r$ is not fixed in \eqref{eq:D6t}. However there is still one piece missing: while imposing the cancellation of the RR tadpole in the compact space ensures vanishing of the RR charge, it is necessary to check that the tension carried by the fluxes and the sources cancels to find a supersymmetric solution. This additional constraint fixes the integer $r$ to be the amount of $F_2$ flux present in the compact space \cite{Marchesano:2006ns}.

\section{Lift to G2 compactifications}
\subsection{General discussion}
Since the IIA flux vacua of concern feature O6/D6 sources, a non-constant dilaton, and $F_2$ flux they lift to pure geometry in 11d. The preservation of minimal supersymmetry further restricts the 7d compactification manifold to  be of G2 holonomy. We now flesh out the details of that, similar to the analysis done in \cite{Kaste:2002xs, Kaste:2003dh,Gaillard:2009kz}.

The 11-dimensional reduction Ansatz to IIA supergravity, in string frame, is given by
\begin{equation}
\label{11dmetric}
\d s_{11}^2 = e^{-\tfrac{2}{3}\phi}\d s_{10}^2 + e^{\tfrac{4}{3}\phi}(\d z + A)^2\,,
\end{equation}
where $z$ runs over the compact M-theory circle: $z = (0,1)$ and locally $F_2=\d A$. Using equations (\ref{10Dmetric}) and (\ref{susy:phi(W)}) the 11-dimensional metric (\ref{11dmetric}) becomes \emph{unwarped}:
\be
\d s_{11}^2 = g_s^{-2/3}\eta_{\mu\nu}\d x^{\mu}\d x^{\nu} + \d s^2_7\,.
\ee
with $\d s_7^2$ the G2-holonomy metric
\be
\label{eq:m7metr}
\d s^2_7 = e^{-2\phi/3}\d s^2_6 + e^{4\phi/3}(\d z + A)^2\,.
\ee

The lift of the $\SU(3)$-structure to a G2 structure can be found in various papers \cite{ Kaste:2002xs, Fayyazuddin:2006gr, Gaillard:2009kz, Schulz:2012wu} and is completely specified by the G2-invariant 3-form $\Phi$\footnote{The normalisation here is chosen to ensure the usual normalisation $\Phi \wedge \star_7 \Phi = 7 d \text{vol}_{\tilde Y}$. This can be easily checked noting that \eqref{eq:m7metr} implies that $d\text{vol}_{\tilde Y} = e^{-4\phi/3 } d\text{vol}_{\tilde X} $ together with the usual normalisations $\frac{i}{8} \Omega \wedge \overline \Omega = \frac{1}{3!} J^3 = d\text{vol}_{\tilde X}$.}
\begin{equation}
 \Phi =e^{-\phi} \Omega_I -  J \wedge (\d z + A) \,,
\end{equation}
We demonstrate the equivalence of the pure spinor equations (\ref{poly}) with $\Phi$'s closure ($\d \Phi = 0$) and co-closure ($\d \star_7 \Phi=0$) in Appendix \ref{App:G2}.

Now let us discuss sources. In the case of vanishing two-form flux, each O6-plane comes with two D6-branes that cancel its charge. The uplift of this setup to the seven-dimensional manifold in M-theory naturally corresponds to a $(\mathbb{R}^3\times S^1)/\mathbb{Z}_2$ singularity, whose appearance suggests non-Abelian gauge theories. Such singularities can be resolved by introducing new cycles, which correspond to the degrees of freedom coming from the D6-branes. The resulting M-theory geometry is known to be a composition of KK-monopoles and Atiyah-Hitchin spaces \cite{Gross:1983hb,Atiyah:1985dv,Gibbons:1986df,Seiberg:1996nz,Seiberg:1996bs}.
In case we have $F_2$ flux the M-theory uplift is not supposed to introduce such degrees of freedom or any singularities. Moreover, $F_2$ flux corresponds to a non-trivial fibration before applying the $\mathbb{Z}_2$ involution on the M-theory side. For this we have to replace the position of the O6-plane by a double cover of Atiyah-Hitchin space. Indeed Atiyah-Hitchin space admits double covers, and we can glue such double covers into the O6-positions. Since the $\mathbb{Z}_2$ involution acts freely on that double cover of Atiyah-Hitchin space, no new singularities and degrees of freedom are introduced.
For this construction to be consistent, we have to ensure that gluing in such a double cover of Atiyah-Hitchin space is consistent with the non-trivial SU(3) structure obeying \eqref{susy:OmegaI}.

Note that this construction of G2 spaces is closely related to the construction obtained by lifting Calabi-Yau orientifolds in which the 7-dimensional manifold has a covering space that corresponds to a direct product of a circle with the Calabi-Yau 3-fold. The essential difference with the construction in this paper is that the covering space is a non-trivial fibre bundle of a $\SU(3)$-structure manifold over a circle, with the twists that defines the fibration encoded in the O6 and D6 sources.  In the case of the CY orientifold the net twist vanishes since the net charges in the sources cancel out. In the case at hand this is not true since the RR tadpole is satisfied by non-trivial flux such that there is a non-trivial twist of the circle over the base. It then turns out that with a non-trivial twist of the circle a G2 space can still be defined on the condition that the $\SU(3)$ holonomy gets relaxed into a more general $\SU(3)$ structure.

\subsection{Moduli spaces}

In general it is a difficult problem to count the moduli of a warped flux compactification since the torsion (or non-closure of $\Omega$ and $J$) complicates the search for fluctuations that leave the pure spinor equations invariant. On top of that the metric on moduli space is not well understood. The ``on-shell'' expression for this metric contains the warp factor, which itself depends on the moduli in a complicated way. The study of these issues is known as \emph{warped effective field theory}, see for instance  \cite{Giddings:2005ff, Shiu:2008ry, Douglas:2008jx, Martucci:2014ska}. For now we observe that our 11d picture at least bypasses the warping issue at first sight. It also facilitates the counting of the moduli since the moduli of G2 compactifications are counted by the third homology class of the G2 space.

The reduction of the $C_3$ potential over the 3-cycles completes the real moduli into complex fields. The complex moduli are then constructed by expanding $\Phi$ and $C_3$ in basis of harmonic 3-forms $\alpha_i$
\be
\Phi + {\rm i} C_3 = \sum_{i=1}^{b^3}\phi_i \alpha^i\,.
\ee
The moduli space metric, determining the kinetic term for the $\phi^i$'s is then derivable from the following K\"ahler potential (see for instance \cite{Beasley:2002db}):
\be
\mathcal{K}  = -3\log\bigg( \frac{1}{14\pi^2} \int_{\tilde Y} \Phi\wedge\star_7\Phi  \bigg) \,.
\ee
It is instructive to write the K\"ahler potential in terms of IIA data
\begin{equation}
\mathcal{K} =
-3 \log \bigg( \frac{1}{12\pi^2} \bigg)
- \log \bigg(\int_{\tilde X}  e^{-4\phi/3} J^3 \bigg)
- 2 \log \bigg(\frac{3i}{4} \int_{\tilde X}  e^{-4\phi/3} \Omega\wedge\bar\Omega   \bigg)
\,,
\end{equation}
where we used \eqref{Omega_J} in the last step (see also \cite{Koerber:2007xk}).
Using \eqref{susy:phi(W)}, it is possible to trade the dilaton $\phi$ for the warped factor $W$ and obtain an expression for the K\"ahler potential which is consistent with the ones suggested in warped effective field theory, that is \footnote{For instance, by redefining $\Phi\to g_2^{2/3}\Phi$ we find that \eqref{Kexplicit} matches 3.38 in \cite{Koerber:2007xk}, where $\mathcal{Z}\sim e^{i\alpha} e^{iJ}$, for a generic phase $\alpha$. We do not care about numerical factors here, since they can be further accommodated by redefinitions of $\Phi$ here or $\mathcal{Z}$ there. Moreover, in the unwarped limit $W\to0$, \eqref{Kexplicit} matches (again, up to some numerical factors) eq.\ 3.48 in \cite{Grimm:2004ua}, which has to be slightly modified by assuming our normalisation \eqref{Omega_J}.}
\begin{align}
\label{Kexplicit}
\calk =
-3 \log \bigg( \frac{g_s^{-4/3}}{12\pi^2} \bigg)
- \log \bigg(\int_{\tilde X} e^{-4W} J^3 \bigg)
- 2 \log \bigg(\frac{3i}{4} \int_{\tilde X}  e^{-4W} \Omega\wedge\bar\Omega   \bigg)
\,.
\end{align}

\subsubsection{Closed string sector}

We first ignore the presence of D6 branes and their associated moduli. This can be achieved by solving the tadpole condition purely with fluxes canceling orientifold charges. We will discuss the effect of D6 branes in the next section.

Using the fibre bundle structure, the most general Ansatz for $\alpha^i$ can be written as
\be \label{def_moduli}
\alpha^i = \Sigma_3^i +\Sigma_2^i\wedge (\d z + A_1)\,,
\ee
where $\Sigma_2$ and $\Sigma_3$ are forms on the 6D base $\tilde{X}$. Note that the $\alpha_i$ must be even under the $\mathbb{Z}_2$ action present in the 7d construction (\ref{Twisted}). Since $\d z + A_1$ is odd we have that $\Sigma_3^i$ is even and $\Sigma_2^i$ is odd.
Closure of $\alpha^i$ implies
\be\label{master}
\d \Sigma_3^i = - \Sigma_2^i\wedge F_2\, , \quad  {\rm and} \quad \d \Sigma_2^i = 0 \, .
\ee
For practical purposes it is easier to phrase the moduli problem in terms of cohomology. The equivalence relation that defines equivalence classes in cohomology consistent with \eqref{master} are
\be \label{equiv}
\Sigma_3^i \sim \Sigma_3^i + \d \lambda_2^i - \lambda_1^i \wedge F_2  \quad  {\rm and} \quad \Sigma_2^i \sim \Sigma_2^i + \d \lambda_1^i \, ,
\ee
for any pair of one- and two-forms $\lambda_{1}$ and $\lambda_{2}$.
From \eqref{master} we see in particular that all even harmonic 3-forms on $\tilde{X}$ provide moduli of the G2 space. It is not difficult to see that in case a mirror IIB flux solutions exists, the K\"ahler moduli of the IIB (conformal) Calabi-Yau space map to the moduli defined by these harmonic 3-forms on the $\SU(3)$-structure base.

We interpret the first equation in \eqref{master} as a differential equation for $\Sigma_3$: If for a given $\Sigma_2$ we can integrate equation (\ref{master}) to find a $\Sigma_3$ then we have found a G2 modulus. For \emph{generic} $F_2$ flux one expects this not to be possible.\footnote{One can integrate (\ref{master}) over a closed 4-surface $\mathcal{C}_4$ that contains a closed 2-surface $\mathcal{C}_2$ encapsulating an O6/D6 source, such as $\mathcal{C}_4 = \mathcal{C}_2 \times \mathcal{D}_2$ then $
\int_{\mathcal{C}_4} \d\Sigma_3 =0 = - Q_{O6/D6}\int_{\mathcal{D}_2}\Sigma_2$,
if we take $\mathcal{D}_2$ the cycle Poincar\'e dual to $\Sigma_2$ then the integral on the RHS can be normalised to +1 such that one indeed finds an obstacle to define $\Sigma_3$.} This is of course a manifestation of moduli-stabilisation due to fluxes and this lift of moduli is T-dual to the lift of complex structure moduli. Similar to the IIB solution it is of course possible that not all such moduli are lifted.

Aside the chiral field moduli that are counted by $b_3(\tilde{Y})$, there can be vector multiplet moduli and they are counted by $b_2(\tilde{Y})$. To compute $b_2(\tilde{Y})$ in terms of data coming from the $\SU(3)$-structure base and data from the fibration (encoded in $F_2$) is completely analogues to the method we used for computing $b_3(\tilde{Y})$. Any even two-form $\gamma^{\alpha}$ can be decomposed as
\be
\gamma^{\alpha} = \Psi_2^{\alpha} +\Psi_1^{\alpha}\wedge (\d z + A_1)\,,
\ee
with $\Psi_2^{\alpha}$ an even two-form on the base and $\Psi_1^{\alpha}$ an odd one-form on the base. Any even $\Psi_2^{\alpha} \in H^2(\tilde{X})$ will lift to an element of $H^2(\tilde{Y})$, but any odd $\Psi_1^{\alpha} \in H^1(\tilde{X})$ can only lift to an element of $H^2(\tilde{Y})$ if there exists an even two-form $\Psi_2^{\alpha}$ such that
\be\label{master2}
\d \Psi_2^{\alpha} =  \Psi_1^{\alpha}\wedge F_2\,, \quad  {\rm and} \quad \d \Psi_1^{\alpha} = 0 \, .
\ee
These forms again satisfy the equivalence relationships
\be \label{equiv2}
\Psi_2^{\alpha} \sim \Psi_2^{\alpha} + \d \kappa_1^{\alpha} - k^{\alpha} F_2  \quad  {\rm and} \quad \Psi_1^{\alpha} \sim \Psi_1^{\alpha} + \d k^{\alpha} \, ,
\ee
for any one-form $\kappa_1^{\alpha}$ and function $k^{\alpha}$.

To summarise, we have found that the moduli space of chiral multiplets of $\tilde{Y}$ is given by $H_3(\tilde{X})$ and possibly a subset of $H_2(\tilde{X})$ if equation (\ref{master}) can be solved. For the sake of phenomenology one is interested in $F_2$ fluxes that do not allow a solution to (\ref{master}). Analogously, the vector multiplets of the moduli space is given by $H_2(\tilde{X})$ and possibly a subset of $H_1(\tilde{X})$ if equation (\ref{master2}) can be solved.

\subsubsection{Open string sector}
The discussion so far allows us to neatly identify the geometric moduli of type IIA models. However this cannot be the whole story for in generic compactifications there will be additional moduli coming from the open string sector, specifically from D6-branes. Since the fibre becomes singular at the position of D6 sources, we cannot simply use the formalism above to count open string moduli and one should be more careful. We have decided not to present a detailed analysis of this since it would obscure the main point of this paper. Furthermore, once fluxes are present one can eliminate open string modes by tuning the amount of flux appropriately. We therefore settle with a brief qualitative discussion.

Two kind of sources will play a role in the following: the first is the Atiyah-Hitchin manifold and the second is the Taub-NUT manifold. It is well known that these two manifolds are uplifts of sources appearing in type IIA, specifically at weak coupling the former becomes an O6-plane and the latter a D6-brane \cite{Gross:1983hb,Atiyah:1985dv,Gibbons:1986df,Seiberg:1996nz,Seiberg:1996bs}. The presence of these objects will distort the six-dimensional geometry and have an impact on the cohomology of the 7-manifold that we are considering.

In type IIA the open string moduli can still be accommodated for in a simple way as long as we only deal with non-intersecting D6 branes, as is the case here. To preserve supersymmetry all sources need to wrap calibrated cycles, and the calibration condition for 3-cycles in Calabi-Yau threefolds gives special Lagrangian cycles \cite{mclean}. The Lagrangian condition amounts to asking that the pullback of the symplectic two form $J$ is zero on the cycles, that is
\be
\imath^* (J) = 0\, ,
\ee
where $\imath$ is the embedding of the brane locus $\tilde S$ into $\tilde X$, and $\imath^*$ is the corresponding pullback to $\tilde S$. While in principle for generic sources this condition might lead to stabilisation of some of the K\"ahler moduli \cite{Marchesano:2014iea} it is always automatically satisfied if all sources are parallel because of the parity of $J$ under the orientifold involution.
The degrees of freedom on the brane are deformations of $\tilde S$, given by sections in the normal bundle of $\tilde S$, and the worldvolume gauge field $A_1$. On special-Lagrangian cycles $J$ defines a map from the normal bundle to the cotangent bundle, mapping a vector field $\xi$ to $\imath^*(J(\xi, \cdot))$ \cite{mclean}. This one-form combines with $A_1$ into a complex field that describes the open string moduli. This completes the story for open string moduli if all sources are parallel because further stabilisation mechanism \cite{Marchesano:2014iea} are absent for the same reason as above. 
While we recalled the calibration condition for the case of branes in Calabi-Yau manifolds it is actually possible to generalise this discussion to the case of manifolds with $SU(3)$ structure that are our case of study. In particular in a supersymmetric background the forms $J$ and $e^{-W} \Omega_R$ are still closed and can be used to calibrate the cycle similarly to the Calabi-Yau case \cite{Koerber:2005qi,Martucci:2005ht,Martucci:2006ij}. Finally one may additionally study the supersymmetric deformations of the calibrated cycles and find that the picture described above does not change when the background geometry is not Calabi-Yau \cite{Koerber:2006hh}.

Now let us understand the counting of open string moduli for $\tilde Y$. The locus $\tilde S$ will be the location of Taub-NUT centres in $\tilde Y$, i.e.\ the locus where the fibre circle degenerates.
We can see that \eqref{def_moduli} only remains well-defined in the presence of D6-branes if $\Sigma_2$ vanishes on $\tilde S$. With parallel sources all forms $\Sigma_2$ odd under the orientifold involution satisfy this criterion and therefore  lift to well defined 3-forms in $\tilde Y$.
Let us now discuss one-cycles on $\tilde S$. If that one-cycle is trivial in $\tilde X$, it is the boundary of a two-chain in the orientifold. The M-theory circle just pinches off at the position of the D6 so that the two-chain times the circle fibration gives a closed three-cycle.
The size of this cycle controls the position of the D6 in M-theory. The Wilson line of $C_3$ on this
cycle gives the Wilson line degree of freedom on the brane.
Thus, it gives a complex modulus of the G2 compactification.
If the one-cycle on $\tilde S$ is not trivial in $\tilde X$, it won't give rise to a two-chain in $\tilde X$, and it does not lead to a non-trivial three-cycle in $\tilde Y$, therefore not adding moduli. Thus, we find that
\be
H^3(\tilde Y) = H^3(\tilde X)_+ \oplus H_{F_2}^2({\tilde X})_- \, ,
\ee
where
\be
H_{F_2}^2({\tilde X}) = \left\{ \Sigma_2 \in H^2({\tilde X})\ |\ \Sigma_2 \wedge F_2 \ {\rm is\ exact}\right\} \, .
\ee
We therefore find agreement between the type IIA and M-theory counting of moduli.

\section{Examples}\label{sec:examples}

In what follows we provide explicit examples. We start by treating one example in detail, including a study of its moduli space, and then we describe a generalisation of this example as a proof of principle that various examples can be found. However, in our generalisation we simply content ourselves with demonstrating the existence, rather than providing a full discussion about features of the models, like moduli spaces, tadpoles et cetera. We proceed by working in the IIA frame and give explicit examples of the supersymmetric Minkowski vacua with just $F_2$ flux and parallel O6/D6 sources since the lift to the G2 manifold was described in detail above.

\subsection{A simple example without flux}\label{sec:simpleexample}

Consider a 6-torus with Cartesian coordinates $x^i$, taking values in the interval $[0,1[$. We assume the following orientifold action:
\begin{align}\label{O6}
\sigma(x) = (-x^1, -x^2, -x^3, x^4, x^5, x^6) \,.
\end{align}
We have 8 O6 planes along directions $x^a$, $a=4,5,6$, and localised in the $x^1,x^2,x^3$. In order to find genuine $\mathcal{N}=1$ supersymmetry (off-shell) we need to further orbifold. The difficulty of adding orbifold involutions is that one typically creates multiple intersecting images of the O6 planes. But as we argued, we do not have a full supergravity description of those and we therefore should make sure the orbifold images create no further intersecting O6 planes. The following $\mathbb{Z}_2^2$ actions
\begin{align}
& \Theta_1(x) = (-x^1, -x^2, x^3, x^4+1/2, -x^5, -x^6+1/2 ) \,, \label{orb1} \\
& \Theta_2(x) = (x^1 , -x^2, -x^3, -x^4, -x^5 +1/2, x^6 +1/2 ) \,, \label{orb2}
\end{align}
do the job since each O6 plane is mapped into itself. The resulting space is not an actual orbifold since the discrete symmetries have no fixed points. As a consequence one can show that the holonomy group is not the full $\SU(3)$ but a restricted $\mathbb{Z}_2^2$ subgroup.

We now follow the method of Kachru-McGreevy to uplift the CY orientifold without fluxes \cite{Kachru:2001je} and find the moduli spectrum. This means we add 16 D6 branes to cancel the tadpole.  We can straightforwardly uplift this setup to a Joyce manifold of type $\mathbb{T}^7/\mathbb{Z}_2^3$, where the $\mathbb{Z}_2^3$ orbifold is generated by
\begin{align}
\alpha(x) = (\Theta_1(x), x^7) \,, \qquad
\beta(x) = (\Theta_2(x), x^7) \,, \qquad
\gamma(x) = (\sigma(x), -x^7) \,.
\end{align}
with $x^7$ the coordinate on the M-theory circle. Note that the resulting 7d space does not have full G2 holonomy but rather a restricted $\SU(2)\times \mathbb{Z}_2^2$. This is a consequence of the freely-acting $\mathbb{Z}_2^2$ involutions (\ref{orb1}, \ref{orb2}). If we lift the 6-torus with the O6 involution (\ref{O6}) and blow up the resulting 7d orbifold we obtain $K3\times \mathbb{T}^3$. If we then add the first $\mathbb{Z}_2$ involution (\ref{orb1}) it is freely acting on $\mathbb{T}^3$ and is Enriques involution on $K3$. This effectively lifts to the Enriques CY $\times S^1$ \cite{Klemm:2005pd}. The second $\mathbb{Z}_2$ involution (\ref{orb2}) then ensures one ends up with a G2 manifold in 11d but with restricted holonomy $\SU(2)\times \mathbb{Z}_2^2$.

We now  compute the spectrum from a 10d and an 11d viewpoint. In 10d, we have to compute the untwisted cohomology first. Then, additional two-/three-forms may appear from resolution of singularities. In this case, however, the orbifold has no fixed points, so there are no singularities and therefore no additional forms. Finally, we have to consider the open string spectrum given by D6's. 

Out of all forms on $\mathbb{T}^6$, there are no one-forms that are invariant under $\langle \Theta_1, \Theta_2\rangle$. The invariant forms are 3 two-forms ($\d x^{16},\d x^{25},\d x^{34}$), and 8 three-forms\footnote{These three-forms are: $\d x^{123},\d x^{124},\d x^{135},\d x^{145},\d x^{356},\d x^{456},\d x^{236},\d x^{246}$.}. Therefore, we have $h^{2,1}=3, h^{1,1}=3$, corresponding respectively to 4 ($h^{2,1}+1$) $\caln=2$ hypermultiplets (considering the dilaton), and 3 $\caln=2$ vector multiplets. By taking into account the orientifold action, we get a $\caln=1$ theory, and all 4 hypermultiplets become chiral multiplets, while the vector multiplets split following the behaviour of the $(1,1)$-forms under $\sigma$. We only have untwisted 2-forms, and these are $\sigma$-odd, giving thus other 3 chiral multiplets. The (untwisted) cohomology therefore accounts for 7 chiral multiplets. The 16 D6 branes needed for tadpole cancellation provide 16 vector multiplets and 48 chiral multiplets. Summing up, the final spectrum is given by 16 vector multiplets and 55 chiral multiplets.

In 11d, we expect to find the same $\caln=1$ spectrum solely from untwisted and twisted cohomology of $\mathbb{T}^7/\mathbb{Z}_2^3$. There are no one- nor two-forms of $\mathbb{T}^7$ which are invariant under $\langle \alpha,\beta,\gamma\rangle$. The only invariant forms are 7 three-forms ($\d x^{124},\d x^{135},\d x^{456},\d x^{263},\d x^{167},\d x^{257},\d x^{347}$). Therefore, the untwisted cohomology yields to 7 chiral multiplets, as in 10d. The spectrum from singularity resolution must then match the open string spectrum found in 10d. For that we have to count the 2-/3-forms from singularity resolution. Elements $\alpha,\beta,\alpha\beta,\alpha\gamma,\beta\gamma,\alpha\beta\gamma$ have no fixed points, and the only element having fixed points is $\gamma$. Its fixed points are 16 $\mathbb{T}^3$'s, parallel and along 456, and these are all distinct from each other under the action of the remaining elements of the orbifold. The local form of the singularities at the fixed $T_3$'s is $\mathbb R^4/\mathbb{Z}_2 \times \mathbb{T}^3$, and resolving each of these as in \cite{Kachru:2001je,Joyce:1} yields a two-form and 3 three-forms. Eventually, we have $b_2=16$ vector multiplets and $b_3=55$ chiral multiplets, as in 10d.

\subsection{With fluxes}\label{sec:simpleexample2}

We expect that the presence of fluxes stabilises some moduli, so that the spectrum gets reduced. A way of thinking is via mirror symmetry: we know that the (supersymmetric) IIB dual model is of DRS (GKP) type \cite{Dasgupta:1999ss} (\cite{Giddings:2001yu}), where fluxes stabilise complex structure moduli. Since complex structure moduli are dual to IIA K\"ahler moduli, we expect some of the latter to be lifted by $F_2$. Also, due to fluxes, the tadpole cancellation is now satisfied with fewer D6's than in the fluxless case, so that the open string spectrum is reduced too.

We operate the moduli counting following the generic discussion developed above, see \eqref{master}. For simplicity we assume $F_2$ is such that the tadpole cancellation is satisfied in absence of D6's.  We twist the torus using the following Maurer-Cartan forms, i.e.\ with\footnote{Notice here $e^a$ correspond to $\tilde e^a$ in \cite{Grana:2006kf}.}
\begin{align}
 e^4 = \d x^4 + N x^1 \d x^2  \,,\\
 e^5 = \d x^5 + N x^1 \d x^3  \,, \\
 e^6 = \d x^6 + N x^2 \d x^3  \,,
\end{align}
and $e^i = \d x^i$ with $i=1,2,3$. The geometric flux $N$ is quantised as $N \in\mathbb{Z}$ since our coordinates run from 0 to 1. One can check that this nilmanifold algebra is consistent under the orbifold and orientifold actions, since
\begin{align}
& \Theta_1(e) = (-e^1, -e^2, e^3, e^4, -e^5, -e^6 ) \,, \\
& \Theta_2(e) = (e^1 , -e^2, -e^3, -e^4, -e^5, e^6 ) \,,\\
& \sigma(e) = (-e^1, -e^2, -e^3, e^4, e^5, e^6) \,,
\end{align}
consistent with the Maurer--Cartan relations. Again, we have 8 O6's, along directions $a=4,5,6$, each one mapped into itself by the orbifold action. On a nilmanifold the cohomology can be computed by restricting to the left-invariant cohomology and one can then verify that the only non-trivial cohomology groups are \cite{Marchesano:2006ns}
\begin{align}
& H^2(\tilde{\mathbb{T}}^6/\mathbb{Z}_2^2,\mathbb{Z}) = \mathbb{Z}^2 \,,\\
& H^3(\tilde{\mathbb{T}}^6/\mathbb{Z}_2^2,\mathbb{Z}) = \mathbb{Z}^6 \times \mathbb{Z}_N  \,,
\end{align}
with $\mathbb{Z}_N$ representing the torsion piece. More explicitly, the closed and non-exact 2-forms surviving the orbifolding are $v_1=e^{16}+e^{25}$ and $v_2=e^{16}-e^{34}$, while the closed and non-exact 3-forms are $e^{124}$, $e^{135}$, $e^{145}$, $e^{236}$, $e^{246}$, $e^{356}$ (and the torsion factor is given by $e^{123}$, since $Ne^{123}$ is exact). A solution to the supersymmetry equations, consistent with the orbifolding, is given by:
\begin{align}
& J=e^{16}-e^{25}-2e^{34}=2v_2-v_1 \,,\\
&g_s F_2^B= N \big[ -\tilde{e}^{16}+\tilde{e}^{25}-4\tilde{e}^{34} \big] \,,\\
&g_s F_2^S=e^{W}\star_3\d e^{-4W}\,,
\end{align}
where the warp factor is sourced by O6's only. We have split the $F_2$ field strength into a background and a sourced part as discussed around equation (\ref{F_2fromOmegaI}). As we discussed, given the K\"ahler deformations $\delta J=t^a \omega_a$, with $\omega_a\in H^2_-(\tilde X)$ ($a=1,2$), the K\"ahler modulus $t^a$ remains massless if there exists a $\gamma_a\in\Omega^3(\tilde X)$ such that
\begin{align}
\d\gamma_a=\omega_a\wedge F_2 \,.
\end{align}
Otherwise, the modulus is lifted. Let us take only $F_2^B$, since one can show that $v_1\wedge F_2^S\sim\d[e^{-3W}(e^{623}+e^{531})], v_2\wedge F_2^S\sim\d[e^{-3W}(e^{623}-e^{412})]$. In other words, the backreaction has no effect on stabilising closed string moduli. We find that only $\omega_1=2v_2-v_1=J$ does satisfy the equation, with trivial 3-form, $\gamma_1=0$. The corresponding modulus, $t_1$, is not stabilised by $F_2$. Not surprisingly, this follows directly from primitivity condition $J\wedge F_2^B=0$. On the other hand, we cannot find any other 3-form satisfying the equation for any choice of $\omega_2\in H^2_-(\tilde X)$, meaning $t_2$ is lifted. Without loss of generality, we take $\omega_2=v_2$. One can check that this is indeed the correct scenario by computing the superpotential (off-shell quantity, here $J=t^a\omega_a$):
\begin{align}
\calw\equiv\int_{\tilde X} J^2 \wedge F_2^B
\sim t_2^2 \int_{\tilde X} w_2^2 \wedge F_2^B
\sim t_2^2
\end{align}
where, due to primitivity $w_1\wedge F_2^B=0$, $t_1$ disappeared. So in total we find $7$ chiral multiplets in our moduli space. Concerning the vector multiplets we potentially pick up a multiplet from the first cohomology class of the base $\tilde{X}$. But the way we orbifolded the twisted torus base, does not allow non-trivial one-cycles and we therefore remain with two vector multiplets.

\subsection{A class of examples}

We now describe a generalisation of the previous example but not in the same amount of detail. We do not discuss the moduli spaces, flux quantisation or the tadpole condition, but it can be done following the steps outlined in the above example.

We start again from groupmanifold covering spaces which are further orbifolded and orientifolded such that we obtain $\mathcal{N}=1$ (off-shell) supersymmetry. In the covering space we can then rely on the existence of a global co-frame build from the Maurer-Cartan forms $e^i$ with $i=1, \dots, 6$, obeying
\begin{equation}\label{MC}
\d e^{i} = -\tfrac{1}{2}f^{i}_{\ j k} e^j \wedge e^k \text{,}
\end{equation}
with $f$ traceless Lie algebra structure constants. To classify solutions that cannot be related by simple relabelling, or coordinate transformations, one needs to fix a basis for the structure constants. For that we  follow the strategy outlined in \cite{Danielsson:2011au}:  one first fixes the $\mathbb{Z}_2$ orientifold and the orbifold group and then consider the possible Lie algebras consistent with the orientifold and orbifold group. Like in our previous example the O6 plane involution $\mathcal R$ is fixed as follows:
\begin{equation}
\label{exorientifold}
\mathcal R :  (e^1,e^2, e^3, e^4, e^5, e^6) \, \rightarrow \left(-e^1,-e^2,-e^3,e^4,e^5,e^6\right)\,,
\end{equation}
and the $\mathbb{Z}_2 \times \mathbb{Z}_2$ orbifold action $\Theta_1, \Theta_2$ are:
\begin{align}
\label{exorbifold1}
\Theta_1 &: (e^1,e^2,e^3,e^4,e^5,e^6) \, \rightarrow (-e^1,-e^2,e^3,e^4,-e^5,-e^6)\,,\\
\label{exorbifold2}
\Theta_2 &:(e^1,e^2,e^3,e^4,e^5,e^6)\, \rightarrow (e^1,-e^2,-e^3,-e^4,-e^5,e^6)\,,
\end{align}
Both sides of \eqref{MC} must transform in the same way under the orbifold and orientifold. Demanding this sets all structure constants to zero except those gathered in the matrix
\begin{equation}
\label{rmatrix}
r =
\begin{bmatrix}
f^{5}_{\ 3 1} &  f^{5}_{\ 4 6} & -f^{2}_{\ 6 3} & -f^{2}_{\ 4 1} \\
-f^{4}_{\ 1 2} & f^{3}_{\ 6 2} & -f^{4}_{\ 5 6} & -f^{3}_{\ 5 1} \\
-f^{6}_{\ 3 2} & f^{1}_{\ 4 2} & f^{1}_{\ 5 3} & f^{6}_{\ 5 4} \text{.}
\end{bmatrix}\,.
\end{equation}

To obtain an interesting solution, at least one of the structure constants must be nonzero. Imposing $f^{5}_{\ 3 1} \neq 0$ and
demanding that the structure constants obey the Jacobi identities restricts them to one of three possible classes
\footnote{This classification corrects a mistake in \cite{Danielsson:2011au} where an extra class was found, labeled C, because a number of additional structure constants were erroneously not set to zero for that class C. Doing so makes that class equal to class B of this paper, after relabelling.}
\begin{align}
\text{Class A:\,\, }r &=
\begin{bmatrix}
f^{5}_{\ 3 1} &  f^{5}_{\ 4 6} & -f^{2}_{\ 6 3} & -f^{2}_{\ 4 1} \\
-f^{4}_{\ 1 2} & \frac{ f^5_{\ 4 6}  f^4_{\ 1 2}}{ f^5_{\ 3 1}} & -\frac{ f^4_{\ 1 2}  f^2_{\ 6 3}}{ f^5_{\ 3 1}} & \frac{ f^2_{\ 4 1}  f^4_{\ 1 2}}{ f^5_{\ 3 1}} \\
-f^{6}_{\ 3 2} & \frac{ f^5_{\ 4 6}  f^6_{\ 3 2}}{ f^5_{\ 3 1}} & \frac{ f^2_{\ 6 3}  f^6_{\ 3 2}}{ f^5_{\ 3 1}} & - \frac{ f^6_{\ 3 2}  f^2_{\ 4 1}}{ f^5_{\ 3 1}}
\end{bmatrix}\,, \\ \nonumber
\text{Class B:\,\, }r&=\begin{bmatrix}
f^{5}_{\ 3 1} & 0 & 0 & -f^{2}_{\ 4 1} \\
-f^{4}_{\ 1 2} & 0 & 0 & -f^{3}_{\ 5 1} \\
0 & 0 & 0 & 0
\end{bmatrix}\,,\qquad\\ \nonumber
\text{Class C:\,\, }r &=
\begin{bmatrix}
f^{5}_{\ 3 1} & 0 & 0 & 0 \\
0 & 0 & 0 & -f^{3}_{\ 5 1} \\
0 & 0 & f^{1}_{\ 5 3} & 0
\end{bmatrix}
\,.
\end{align}

We can now look at the way the conditions on $J$ and $\Omega$ constrain our groupmanifold. To perform the classification of possible groupmanifolds, we work in the smeared limit where the warp factors are set to constant. They can eventually be inserted for the case one wishes to study more closely after the classification has been completed.
Because $J$ and $\Omega_R$ must be odd under the orientifold and invariant under the orbifold, their form is restricted to
\begin{align}
J &= k^1 e^{52} + k^2 e^{4 3} + k^3 e^{6 1}\text{,}\\
\Omega_R &= \mathcal{F}_1 e^{312} + \mathcal{F}_2 e^{462} + \mathcal{F}_3 e^{563} + \mathcal{F}_4 e^{541} \text{.}
\end{align}
One sees that \eqref{susy:OmegaR} is immediately satisfied and $\Omega_R$ imposes no constraints on the algebra of our groupmanifold. It is still possible that $\Omega_I$ imposes constraints on the groupmanifold algebra via \eqref{susy:OmegaI}. However, rather than seeing this equation as a constraint on the groupmanifold, we can see it as telling us to which background value we should set  $F_2$ to realise the groupmanifold we choose.
This still leaves us with the constraints on the structure constants imposed by $J$ via \eqref{susy:J}. For class A, we find
\begin{equation}
\label{classAconstraint}
\begin{split}
f^5_{\ 4 6}  \big[ k_1   f^5_{\ 3 1} - k_2   f^4_{\ 1 2} + k_3 f^6_{\ 3 2} \big] &= 0 \,, \\
 k_1 f^5_{\ 3 1} + k_2 f^4_{\ 1 2} - k_3 f^6_{\ 3 2} &= 0 \,, \\
f^2_{\ 6 3} \big[ k_1  f^5_{\ 3 1}  - k_2 f^4_{\ 1 2}   - k_3   f^6_{\ 3 2} \big] &= 0 \,, \\
 f^2_{\ 4 1}\big[ k_1  f^5_{\ 3 1} + k_2  f^4_{\ 1 2} + k_3 f^6_{\ 3 2}   \big] &= 0 \,,
\end{split}
\end{equation}
must be satisfied, while for class B we get
\begin{equation}
\label{classBconstraint}
\begin{split}
k_1 f^5_{\ 1 3} + k_2 f^4_{\ 1 2} = 0\,, \\
k_1 f^2_{\ 4 1} -  k_2 f^3_{\ 5 1} = 0\,,
\end{split}
\end{equation}
with all structure constants not appearing in these equations being required to be zero. The constraints on class C do not allow nontrivial solutions. In the end we find four possible types of group manifolds covering spaces, shown in table \ref{tableexamples}, and manifolds obtained  from these by the simultaneous exchange of the axes $4 \leftrightarrow 6$ and $1 \leftrightarrow 3$

\begin{table}[h]
\center
\begin{tabular}{l|l}
Type         & Nonzero f's \\ \hline
Nilmanifold  & $f^5_{\ 1 3}$, $f^4_{\ 1 2}$  \\ \hline
Nilmanifold  & $f^5_{\ 1 3}$, $f^4_{\ 1 2}$, $f^6_{\ 3 2}$ \\ \hline
Solvmanifold & $f^5_{\ 1 3}$, $f^2_{\ 6 3}$, $f^6_{\ 3 2}$\\ \hline
Solvmanifold & $f^5_{\ 1 3}$, $f^3_{\ 5 1}$, $f^4_{\ 1 2}$, $f^2_{\ 4 1}$
\end{tabular}
\caption{Groupmanifolds satisfying the constraints imposed on our class of examples. \label{tableexamples}}
\end{table}

Relationships between the structure constants and the factors $k^i$ of $J$ are determined by either \eqref{classAconstraint} or \eqref{classBconstraint}.
All these algebras can be implemented by compact manifolds as discussed for instance in in the classification of \cite{Grana:2006kf}, where they are referred to as n4.6, n3.5, s3.3 and s2.5 respectively\footnote{Strictly, our solvmanifolds are of a larger class that contain s3.3 and s2.5 within them and compactness has only been fully established for s3.3 and a subset of  s2.5 \cite{Grana:2006kf}.}.

If we wish to make sure that no intersecting O-planes appear, we must make sure that $\Theta_1$ and $\Theta_2$ do not create any additional O-planes. So far, we have only discussed how $\Theta_1$ and $\Theta_2$ act on the Maurer-Cartan forms and we still have some freedom in choosing how they act on the coordinates, so long as this does not alter how they act on the Mauer-Cartan forms. By making sure $\Theta_1$, $\Theta_2$ and all other involutions except for $\mathcal{R}$ are freely acting on at least one direction by making them act as $x_i \rightarrow x_i + 1/2$, we ensure they do not create additional O-planes. This is straightforward to implement for the nilmanifolds as demonstrated for a specific case in section  \ref{sec:simpleexample}. However, we were unable to do this for the solvmanifolds. The problem is that if e.g.
\begin{equation}
e^5 = dx^5 + x^3 dx^1 \,,
\end{equation}
then a shift of $x^3$ does not leave $e^5$ invariant; making it disallowed. Due to restrictions like this, it is impossible to implement enough shifts to make all the required involutions freely acting.

In conclusion, we find two types of nilmanifolds and two types of solvmanifolds that are compatible with our demands, which can be seen in table \ref{tableexamples}. However, we could not find a way to avoid the appearance of intersecting O-planes for the solvmanifolds. The presence of intersecting O-planes may be problematic. The remaining set of examples is a handful since one has some freedom to take an integer choice of metric flux and $F_2$ flux to cancel the RR tadpole for 8 O6 planes. Allowing D6 branes as well, the size of this set enlarges from order 10 to roughly order 100.

\section{Comments on supersymmetry breaking}

\subsection{SUSY-breaking fluxes and G2 structures }
There is a rather straightforward way to break supersymmetry of the IIA flux backgrounds and maintain reasonable control. 
The inspiration comes from the supersymmetry-breaking on the mirror IIB side that occurs whenever fluxes are imaginary self-dual (ISD)  but with certain components that are not $(2,1)$ and primitive \cite{Giddings:2001yu}. There exist simple solutions of that kind that can be T-dualised to IIA Minkowski solutions with just $F_2$ flux. The simplest example is perhaps the following.

Consider (an orbifold of) the 6-torus with coordinates $y^1,\ldots, y^6$. Assume the following form of the ISD flux:
\begin{equation}
F_3 = f \d y^4\wedge \d y^5\wedge \d y^6 \,, \quad H_3 = e^{\phi_0}\star_6 F_3 \,.
\end{equation}
This solution is such that the ISD three-form flux on the six torus has a $(0,3)$ component  \cite{Blaback:2013taa} and hence breaks SUSY. If we now T-dualise 3 times along $y^1, y^5, y^6$  we obtain the following IIA twisted torus background (ignoring warping)
\begin{align}
& \d s^2 =\d s^2_4 + (\d y^5)^2 + (\d y^6)^2 + (e^1)^2 + (\d y^2)^2 + (\d y^3)^2 + (\d y^4)^2\,, \\
& F_2 = - f \d y^4 \wedge e^1 \,,
\end{align}
where the one-form $e^1$ is twisted as follows:
\begin{align}
 e^1 = d y^1 + f y_2 e^{\phi_0} d y^3\,.
\end{align}
The O6 stretches along the T-duality directions $e^1, y^5$ and $y^6$ such that its $\mathbb{Z}_2$ involution on the internal space works as $y^{2,3,4}\rightarrow -y^{2,3,4}$. Depending on the choice of lattice on the resulting Nilmanifold this action has a number of invariant 3-dimensional subspaces such that the number of O6 planes is fixed and so is the total tension $T_6$. The number $f$ can be called metric flux and relates to the O6 tension $T_6$ as follows
\be
T_6 = f^2 e^{-\phi_0}\,.
\ee
If this solution is lifted to 11d, one obtains again a solution to the 11D Einstein equations with an unwarped 4D Minkowski piece and a flat compact 7d internal manifold. Since supersymmetry is not preserved it cannot have G2 holonomy. Instead it is not difficult to argue one obtains a Ricci-flat G2 \emph{structure}. So the IIB method for breaking supersymmetry and lifting some moduli by fluxes has turned into a purely geometric picture in which G2 holonomy manifolds are altered into Ricci-flat G2 \emph{structures} that break supersymmetry and with typically less moduli than their G2 holonomy counterparts\footnote{This is not unrelated to the ideas presented in \cite{DallAgata:2005zlf, Danielsson:2014ria}.}.

\subsection{A non-homogenous extension of the Scherk-Schwarz mechanism}

We want to emphasise that this is a very close cousin of the  mechanism found by Scherk and Schwarz (SS) to stabilise part of the torus moduli in 10- or 11-dimensional supergravity and break supersymmetry \cite{Scherk:1978ta, Scherk:1979zr}. This is achieved by dimensionally reducing over a twisted torus with a flat metric. The deviations away from the flat metric are massive and together with the massless moduli the degrees of freedom are captured by a gauged supergravity of the no-scale type that has a Minkowski vacuum with all supersymmetry broken. To be more precise, the breaking of supersymmetry and the lifting of torus moduli depends in a subtle way on the quantisation rules for the structure constants \cite{DallAgata:2005zlf, Grana:2013ila} invisible in the gauged supergravity description. The gauged supergravity itself has the same amount of supersymmetry as the higher-dimensional theory one starts off with.

If one compares the SS mechanism with the more modern approach to moduli stabilisation and supersymmetry-breaking that uses fluxes and orientifolds there are some advantages and disadvantages. An advantage is that we do not have to rely on singular objects that source warping effects whose effective field theory description is more challenging to work out. One is furthermore not constrained by tadpole conditions that bound the size of fluxes since the `metric fluxes' in SS compactifications can be taken to be arbitrarily large (aside the quantisation subtleties previously mentioned). A disadvantage is that SS compactifications of 10/11-dimensional supergravity lead to maximal or half-maximal gauged supergravity descriptions where supersymmetry is broken spontaneously. This is related to the fact that twisted tori are locally homogeneous, which furthermore also strongly restricts the number of examples or `metric fluxes' that can lead to flat metrics. In fact they can easily be classified into a handful of examples \cite{Grana:2013ila}.

Our picture connects the SS mechanism for moduli stabilisation with that of orientifolds and fluxes by demonstrating that there exists a T-duality frame (IIA with O6 planes and $F_2$ fluxes) in which the orientifold lifts to a generalised Scherk--Schwarz space, which differs from the twisted tori by being locally non-homogeneous in 11d. The breaking of supersymmetry in the IIA side can be understood as a T-dual of supersymmetry breaking ISD three-form fluxes in IIB.  This demonstrates that there exists a class of supersymmetry-breaking flat manifolds that lead to  classical Minkowski vacua beyond the purely homogenous examples of Scherk and Schwarz.

\subsection{Non-perturbative quantum effects and uplifting?}

Finally we want to comment further on moduli-stabilisation. The mirror (and 11D lift) of the K\"ahler moduli in IIB remain massless at the classical level. Hence one could speculate on (quantum)-corrections to the 11D supergravity action that can further stabilise those. In the IIA frame some mechanism has been suggested in \cite{Palti:2008mg} that is supposed to mirror the Large Volume Scenario (LVS) \cite{Balasubramanian:2005zx} in IIB.
Here we briefly discuss possible IIA mirror of KKLT \cite{Kachru:2003aw} and its M-theory lift. 
We hope that the geometric picture in M-theory 
could potentially elucidate some of the questions raised about the consistency of moduli stabilisation \cite{Sethi:2017phn}.

The natural quantum effects to consider are Euclidean D2 branes wrapping calibrated cycles in the IIA frame or gaugino condensation on the reduced 7D YM theory living on wrapped D6 branes. These quantum effects can potentially lift the massless modes and turn the non-SUSY Minkowski vacuum into a SUSY AdS vacuum. We expect that, in the 11D picture, these two effects are either corresponding to Euclidean M2 branes or to localised gravitino condensation in the Ricci-flat G2 structure. It would be interesting to study this in some depth.

Then one can further speculate about possible uplift sources that break supersymmetry of the resulting AdS vacuum. The natural candidates are the mirrors to anti-D3 branes, which are anti-D6 branes \cite{Palti:2008mg}. For them to be consistent one should be able to redshift their tensions through local throats which should be mirrors to the Klebanov-Strassler throats. They have been reported to exist in \cite{Gaillard:2009kz}. This uplift method, from a 11d viewpoint is entirely geometrical since the anti-D6 branes lift to local anti-Taub-NUT spaces. We therefore suspect that the consistency of anti-D3 uplifting to dS \cite{Moritz:2017xto, Gautason:2018gln} can be more easily analysed in this framework since most of the complicated flux terms in these computations can all be reshuffled into the 11d curvature term. We hope to come back to this in the future.

\section{Discussion}
In this paper we have argued that a class of supersymmetric IIA flux compactifications to 4d $\mathcal{N}=1$ Minkowski vacua are in one-to-one correspondence with compactifications of 11-dimensional supergravity on new G2 spaces. The IIA flux compactifications for which this works are constrained by demanding that the only sources are O6 planes and D6 branes and the only fluxes are  RR $F_2$ fluxes. Under these conditions the uplift to 11 dimensions is guaranteed to be completely geometrical, in other words IIA vacua then lift to unwarped 4d Minkowski space times a 7d Ricci flat geometry. The requirement of minimal supersymmetry then further constrains the 7d geometries to have G2 holonomy. 

It was well-known that in the fluxless limit the IIA orientifolds lift to G2 spaces and in the case of toroidal orientifolds there seems to be a nice connection with the Joyce class \cite{Kachru:2001je}. Here we point out that the general structure of the supersymmetric compactifications of IIA with the specified restrictions are sufficiently well-understood in general terms \cite{Grana:2005jc, Grana:2006kf} in order to analyse the uplift to 11d. Especially in the case where the sources are all parallel to each other, the resulting geometries are under control since it is known how to describe the sources in a localised manner. The geometries are specific $\SU(3)$-structure manifolds which we denoted $\tilde{X}$. The  non-Calabi-Yau nature of $\tilde{X}$  is a consequence of the $F_2$ flux backreaction. The $F_2$ flux has furthermore a non-trivial interplay with the  topology such that the RR tadpole can be satisfied without having to cancel all O6 charges with D6 charges \cite{Marchesano:2006ns}. The uplifted geometry, denoted $\tilde{Y}$, is then a twisted circle fibration of $\tilde{X}$ modded out by the $\mathbb{Z}_2$ action of the O6 planes (\ref{Twisted}):
\begin{equation}
\tilde{Y} = \frac{\tilde{X} \ltimes S^1}{\mathbb{Z}_2}\,.
\end{equation}
The CY limit (obtained by cancelling all O6 charges with D6 charges) leaves no room for non-trivial $F_2$ fluxes and we obtain an untwisted circle fibration. 

We have given a general discussion of moduli counting both in IIA and in 11-dimensional supergravity and argued that the moduli counting matches. For the sake of being explicit and as a proof of existence we have presented a handful of examples by finding explicit $\SU(3)$-structures based on orbifolds of twisted tori. The minor drawback of that approach is that the resulting holonomy is not the full G2 group but a non-trivial subgroup thereof. We expect that the recent developments in constructing non-trivial $\SU(3)$-structures \cite{Larfors:2010wb, Larfors:2018nce} should be very relevant in this context. So if more progress on the construction of explicit $\SU(3)$-structures is to occur, then our construction implies immediately a new class of G2 spaces.

Our main interest, however, was not only to enlarge the class of G2 spaces. Instead we consider it very interesting that a class of warped flux compactifications in IIA, which are mirrors of IIB flux compactifications with 3-form fluxes, are completely geometrised. As a consequence the moduli space of a warped compactification, which is the topic of warped effective field theory (WEFT), can be  rephrased in terms of ordinary Kaluza-Klein theory of 11d supergravity on G2 spaces. We hope this will elucidate some of the difficulties in understanding WEFT, which would be interesting to address in the future. Finally, we argued that this geometrisation of IIA flux compactifications should be useful in investigating recent concerns\footnote{See references \cite{Kachru:2018aqn, Cicoli:2018kdo} for complementary viewpoints that refute the concerns.} about non-perturbative moduli-stabilisation \cite{Sethi:2017phn} and the supersymmetry-breaking dS uplifts on top of it \cite{Moritz:2017xto, Gautason:2018gln}, which we consider interesting directions for future research.

\acknowledgments The work of GS is supported in part by the DOE grant DE-SC0017647 and the Kellett Award of the University of Wisconsin. 
The work of H.T. was supported by the EPSRC Programme Grant EP/K034456/1. The work of GV is  supported  by  the  European  Research  Council grant ERC-2013-CoG 616732 HoloQosmos. The work of TVR is supported by the FWO odysseus grant G.0.E52.14N and the C16/16/005 grant of the KULeuven.  We acknowledge support from the European Science Foundation HoloGrav Network, the Belgian Federal Science Policy Office through the Inter-University Attraction Pole P7/37, and the COST Action MP1210 `The String Theory Universe'. The work of GZ was supported by  the HKRGC grants HUKST4/CRF/13G
and 1630441 and the NSF CAREER grant PHY-1756996.
\newpage

\appendix

\section{Conventions}\label{App:conventions}
Our form conventions are:
\begin{align}
& \alpha_p = \frac{1}{p!} \alpha_{m_1 \ldots m_p} \d x^{m_1\ldots m_p}\,,\\
& \omega_q \lrcorner \alpha_p = \frac{1}{q (p-q)!} \omega^{m_1 \ldots m_q} \alpha_{m_1 \ldots m_q\ldots m_p} \d x^{m_{q+1}\ldots m_p}\,.
\end{align}
Concerning $\SU(3)$-structures we follow the notation and conventions of  \cite{Grana:2006kf}. A summary of the most relevant equations for $\SU(3)$ structures, that fix our conventions are:
\begin{align}
\label{Omega_J}
& \frac{3i}{4} \Omega \wedge \bar \Omega =  J^3 \, , \\
\label{astOmega}
& \star_6 \Omega =  i \Omega\,,\\
& \star_6 J = \tfrac{1}{2} J\wedge J\,,\\
& J\wedge J\wedge J = 6 \d \text{Vol}\,,\\
\label{torsion:dJ}
& \star_6 W_2 =  - J \wedge W_2\,.
\end{align}
The convention for the Hodge-$\star$ is the one in footnote 13 of \cite{Grana:2006kf}. This means that, given a 2-form $w_2 = w^{(2,0)} + w^{(0,2)} + w^{(1,1)}$:
\begin{align}
& \star_6 w^{(2,0)} = J \wedge w^{(2,0)} \,, \\
& \star_6 w^{(0,2)} = J \wedge w^{(0,2)} \,, \\
& \star_6 w^{(1,1)}_{prim} = - J \wedge w^{(1,1)}_{prim} \,,
\end{align}
where the last identity is valid only for the primitive part (if any) of $w^{(1,1)}$, i.e.\ such that $J^{mn} w_{mn}^{(1,1)prim} = 0$ (or $J\wedge J \wedge w^{(1,1)}_{prim}$).

\section{Localised O6 solution}\label{App:localised}
The solution with localised sources is straightforward to obtain by dressing forms with the appropriate warping as explained in \cite{Grana:2006kf, Blaback:2010sj}. A factor of $e^W$ ($e^{-W}$) is added in $J,\Omega$ for every coordinate one-form $e^\alpha_+$ ($e^\alpha_-$) parallel (transverse) to O6's (we use a tilde to refer to solutions in the smeared case):
\begin{equation}
e^\alpha_+ = e^W \tilde e^\alpha_+			 \,,\qquad
e^i_- = e^{-W} \tilde e^i_-				\,.
\end{equation}
This method only works for localising sources that are all parallel.

We can now compute $F_2$ from \eqref{susy:OmegaI} and this requires $\Omega_I$. The anti-holomorphic O6 involution $\sigma$ is supposed to act on the pure spinors as \cite{Grana:2006kf}:
\begin{equation}
\sigma(J) = - J \,,\qquad
\sigma(\Omega) = - \bar{\Omega} \,.
\end{equation}
From this fact we find that $J$ has a foot along and the other transverse to the O6's. $\Omega_R$ can have either a foot transverse and 2 feet parallel to O6's or all 3 feet transverse. On the other hand, $\Omega_I$ can have either a foot parallel and 2 transverse or all parallel. In our case, $\Omega_I = -2 e^{124} + e^{135} + e^{236} + 2 e^{456} = -2 e^{-W} \tilde e^{124} + e^{-W} \tilde e^{135} + e^{-W} \tilde e^{236} + 2 e^{3W} \tilde e^{456}$, and then
\begin{align}
\d ( e^W \Omega_I) & = \d (-2 \tilde e^{124} + \tilde e^{135} +  \tilde e^{236} + 2 e^{4W} \tilde e^{456}) 	\\
& = [ 2 \d e^{4W} \wedge \tilde e^{456} + 2 e^{4W} (\tilde e^{1256} - \tilde e^{1346} + \tilde e^{2345}) ]\,.
\end{align}
In order to compute $F_2$ one needs to compute the Hodge-$\star_6$ with respect to the (warped) metric $g$:
\begin{align}
\star_6 \tilde e^{abcd} & = \frac{1}{2} \sqrt{g} g^{aa'} \cdots g^{dd'} \epsilon_{efa'...d'} \tilde e^{ef}	\\
& = \frac{1}{2} \sqrt{\tilde g} g^{aa'} \cdots g^{dd'} \epsilon_{efa'...d'} \tilde e^{ef}	\,,
\end{align}
where $g^{aa'} = e^{\mp 2W} \tilde g^{aa'} $ if $a,a$ are directions parallel or transverse to O6's respectively. The unwarped metric $\tilde g_{mn}$ can be extracted from the expression for $J$, since $J = \tilde J =  i \tilde g_{i \bar\jmath} \tilde z^i \wedge \bar{\tilde z}^j$. One finds $\tilde g_{i \bar\jmath} = \frac{1}{2} \delta_{i \bar\jmath}$, which in real coordinates means
\begin{equation}
\label{metric}
\tilde g = \diag (1,1,1,4,1,1)			\,.
\end{equation}
Therefore
\begin{equation}
\star_6 \tilde e^{1256}  =		2 \tilde e^{34}	\,,			\qquad
 \star_6 \tilde e^{1346} = 	 \frac{1}{2}	\tilde e^{25}				\,,		\qquad
 \star_6 \tilde e^{2345} = 	\frac{1}{2}		\tilde e^{16}		\,.
\end{equation}
We can finally compute $F_2$:
\begin{align}
g_s F_2
& = - e^{-4W}  \star_6 \d (e^W \Omega_I)							\\
& = 2 e^{4W} \star_6 (\d e^{-4W} \wedge \tilde e^{456} ) 		- 2 ( 2 \tilde e^{34}  -  \frac{1}{2}	\tilde e^{25}	+  \frac{1}{2}	\tilde e^{16}	)			\\
& = e^W \star_3 \d e^{-4W} 			-   4 \tilde e^{34}  + 	\tilde e^{25}	- 	\tilde e^{16}		
\,,
\end{align}
where we have used the fact that $\star_6$ decomposes as $\star_3 \star_3^{\parallel}$ and $\star_3^{\parallel} \tilde e^{456} = \frac{e^{-3W}}{2}$. As we pointed out below \eqref{F_2fromOmegaI}, $F_2$ splits into the background component, plus a component sourced by local objects.
By taking the exterior derivative:
\begin{align}
g_s \d F_2
& = \d( e^W \star_3 \d e^{-4W} ) 			+ 6  \tilde e^{123}  				\\
& = (\tilde \nabla_-^2 e^{-4W} + 6 ) \tilde e^{123}  					\,,
\end{align}
and the Bianchi identity becomes an equation for the warp factor:
\begin{equation}
(\tilde \nabla_-^2 e^{-4W} + 6 ) \tilde e^{123}  = g_s Q \delta \,.
\end{equation}

\section{$\SU(3)$ torsion classes}\label{App:torsion}
The deviation of an $\SU(3)$-structure manifold from having $\SU(3)$ holonomy is measured by the torsion (classes), which expresses the non-closure of $\Omega$ and $J$ via
\begin{align}
& \d J =  \tfrac{3}{2}\Im(\bar{W}_1\Omega) + W_4\wedge J + W_3\,,\\
\label{torsion:dOmega}
& \d\Omega = W_1 J^2 + W_2\wedge J + \bar{W}_5 \wedge \Omega\,.
\end{align}
To gain more understanding of the flux solutions in IIA supergravity we compute the torsion classes. The closure of $J$  implies
\begin{equation}
\label{W1,W3,W4}
W_1 = W_3 = W_4 = 0 \,.
\end{equation}
Since $\d \Omega =  (\d \Omega)^{3,1} +  (\d \Omega)^{2,2} $ (this follows from the fact that $\Omega$ is a complex $(3,0)$-form), and  $\star_6 F_2  = (\star_6 F_2)^{3,1} + (\star_6 F_2)^{2,2} + (\star_6 F_2)^{1,3} $, we can rewrite  \eqref{susy:OmegaR}, \eqref{susy:OmegaI} as

\begin{align}
\label{susy:dOmega3,1}
& (\d \Omega)^{3,1} = (\d W)^{0,1} \wedge \Omega  = - i e^\phi J \wedge F^{2,0}		 \,,		\\
\label{susy:dOmega1,3_dW}
&(\d \bar\Omega)^{1,3} = (\d W)^{1,0} \wedge \bar\Omega  = i e^\phi J \wedge F^{0,2} 	\,,\\
\label{susy:dOmega2,2}
& (\d \Omega)^{2,2} = - (\d \bar\Omega)^{2,2}  = - i e^\phi (\star_6 F_2)^{2,2} 			\,.
\end{align}
By matching these equations with \eqref{torsion:dOmega}, which can be decomposed in
\begin{align}
& (\d \Omega)^{2,2} =  W_2\wedge J = - \star_6 W_2 \,, \\
& (\d \Omega)^{3,1} =  \bar{W}_5 \wedge \Omega \,,
\end{align}
one can extract $W_2$ and $W_5$. From the first one and \eqref{susy:dOmega2,2}, one can immediately read $W_2$:
\begin{equation}
\label{W2 wedge J}
W_2 = i e^\phi  F^{1,1}
\end{equation}
which means that $F^{1,1}$ has to be primitive. Indeed, one can check it from $J \wedge (\d \Omega)^{2,2} = J^2 \wedge W_2 = 0$, which is also $J \wedge (\d \Omega)^{2,2} = -i e^\phi J \wedge\star_6 F^{(1,1)} \sim J^{mn} F^{(1,1)}_{mn} $. Hence, $J^{mn} F^{(1,1)}_{mn} = 0$.

Let us compute $W_5$. By contracting \eqref{torsion:dOmega} with with $\bar\Omega$:
\begin{equation}
\begin{aligned}
\label{W5def}
\bar\Omega \lrcorner \d \Omega
= \bar\Omega \lrcorner (\d \Omega)^{3,1}  				
& =  \bar\Omega \lrcorner (\bar W_5 \wedge \Omega) 			\\
& =	\frac{1}{3!} \bar\Omega^{mnp} \left( 3 \bar W_{5m} \Omega_{npq} -  \bar W_{5q} \Omega_{mnp} \right) \d x^q	 \\
& =	- \frac{1}{3!} \bar\Omega^{ijk} \bar W_{5 \bar\imath} \Omega_{ijk} \ \d \bar z^{\bar\imath}		 \\
& =	- |\Omega|^2 \bar W_5		 \,.
\end{aligned}
\end{equation}
Upon using the supersymmetry equation \eqref{susy:dOmega3,1}, one finds
\begin{equation}
\begin{aligned}
\bar\Omega \lrcorner (\d \Omega)^{3,1}
& = - i e^\phi \bar \Omega \lrcorner  (J \wedge F^{2,0})			\\
& = -  \frac{i e^\phi}{3!} \bar \Omega^{ijk} (J \wedge F^{2,0})_{ijk \bar k}	\d \bar z^k \,,
\end{aligned}
\end{equation}
and finally, since in our conventions $|\Omega|^2 = 8$:
\begin{equation}
\label{W5}
\bar W_5 =
\frac{1}{8} i e^\phi \bar \Omega \lrcorner  (J \wedge F^{2,0})
= \frac{i e^\phi}{3! 8} \bar \Omega^{ijk} (J \wedge F^{2,0})_{ijk \bar k}	\d \bar z^k 		\,.
\end{equation}
For future reference, notice that \eqref{susy:dOmega1,3_dW} implies (by expressing $(\d\bar\Omega)^{1,3}$ in terms of torsion structures):
\begin{equation}
\label{dW and W5}
\d W = W_5 + \bar W_5 \,.
\end{equation}

\section{(Co-)Closure of $\Phi$}\label{App:G2}

Here, we show that the G2-form
\begin{equation}
\Phi = e^{-\phi} \Omega_I - J \wedge (\d z + A)
\end{equation}
is closed and co-closed, iff supersymmetric equations are satisfied. Closure, $\d\Phi=0$, is achieved first by observing that, due to the supersymmetric condition $\d J=0$,
\begin{align}
\d\Phi=\d(e^{-\phi} \Omega_I) -  J \wedge F_2 \,,
\end{align}
Then, by using \eqref{susy:OmegaI}, \eqref{susy:dOmega1,3_dW},\eqref{susy:dOmega3,1}, \eqref{dW and W5}, \eqref{susy:phi(W)}, it is possible to show that the RHS vanishes. In fact:
\begin{align}
\label{e-3WOmega}
\d (e^{-\phi} \Omega_I)
& = - \frac{4}{3} e^{-\phi} \d \phi \wedge \Omega_I - \star_6 (F^{(2,0)} + F^{(0,2)} + F_{prim}^{(1,1)}) \\
& = 2i e^{-\phi} (\bar W_5 \wedge \Omega - W_5 \wedge \bar\Omega ) -  J \wedge (F^{(2,0)} + F^{(0,2)} - F_{prim}^{(1,1)}) 		\\
& = 2  (J \wedge F^{(2,0)} + J \wedge F^{(0,2)} ) -  J \wedge (F^{(2,0)} + F^{(0,2)} - F_{prim}^{(1,1)}) 		\\
& = J \wedge F_2
\,.
\end{align}
In order to prove co-closure, $\d \star _7\Phi=0$, we have to compute $\star_7\Phi$ first. In order to act with the Hodge-$\star_7$, we have to recast $\Phi$ in terms of 7d quantities. We will use siebenbeins. Since the metric is $\d s_{11}^2 = e^{-\tfrac{2}{3}\phi}\d s_{10}^2 + e^{\tfrac{4}{3}\phi}(\d z + A)^2$, seibenbeins are defined as $e^A = (e^a_{(7)}, e^7) = (e^{-\phi/3} e^a, e^{2\phi/3} (\d z + A))$ (while sechsbeins are $e^a$ with $a=1...6$). We can now compute the $\star_7\Phi$ by defining $\epsilon_{abcdef7}\equiv\epsilon_{abcdef}$.
The first terms yields
\begin{align}
\star_7 [ e^{-\phi} \Omega_I ] & = \frac{1}{3!} e^{-\phi} \star_7 [ e^\phi (\Omega_I)_{abc} e^{abc}_{(7)} ]    \\
& = - \frac{1}{3!}   (\Omega_I)_{abc} \frac{1}{3!} \epsilon_{def}^{\ \ \ abc} e^{-\phi/3}  e^{def} \wedge (\d z + A)   \\
& = -  e^{-\phi/3}  \star_6 \Omega_I  \wedge (\d z + A) \\
& = - e^{-\phi/3}   \Omega_R \wedge (\d z + A) \,,
\end{align}
where we used the definition of $\star_6$ and the property $\star_6 \Omega_I=\Omega_R$. The second term is
\begin{align}
\star_7 [ J \wedge (\d z + A)] & = \star_7 [ J_{(7)} \wedge e^7] \\
& = \frac{1}{2 \cdot 4!} J _{ab} \epsilon_{cdef}^{\ \ \ ab}  e^{-4\phi/3} \tilde e^{cdef}  \\
& =  e^{-4\phi/3} \star_6 J   \\
& = \frac{1}{2} e^{-4\phi/3} J \wedge J  \,,
\end{align}
where we used $\star_6 J=\frac{1}{2} J^2$ in the last step.
We can show now that
\begin{align}
\star_7 \Phi & =  - e^{-\phi/3}  \Omega_R \wedge (\d z + A) -   \frac{1}{2} e^{-4\phi/3} J^2
\end{align}
is closed:
\begin{equation}
\d \star_7 \Phi =e^{-\phi/3}   \Omega_R \wedge F_2 -  \frac{1}{2}  \d e^{-4\phi/3} \wedge J \wedge J = 0 \,,
\end{equation}
where we used the supersymmetry equations \eqref{susy:J}-\eqref{susy:OmegaR} in the first step and
\begin{align}
\label{relation_for_coclosure}
- \frac{1}{2}  \d e^{-4\phi/3} \wedge J \wedge J & = \frac{2}{3} e^{-4\phi/3} \d \phi \wedge J \wedge J
=  - e^{-\phi/3} 	F_2 \wedge \Omega_R	
\end{align}
in the final step. The relation \eqref{relation_for_coclosure} can be proven by using (as one may check from \eqref{dW and W5}, \eqref{W5} and some little algebra \footnote{Explicitly,
$\bar W_5 \wedge J^2
= i \frac{e^\phi}{3!8} \bar\Omega^{ijk} (J \wedge F^{(2,0)})_{ijk\bar k} \d\bar z^k \wedge J^2
= - \frac{e^\phi}{16} \bar f \epsilon_{\overline{kbc}} g^{\bar b j}g^{\bar c k} F_{jk} g_{l \bar m} g_{p \bar n} \epsilon^{\overline{kmn}} \d z^{\overline{123}}\wedge\d z^{lp}
= ... = - \frac{e^\phi}{4} \bar f \d z^{\overline{123}} \wedge F_2
= - \frac{e^\phi}{4} \bar\Omega \wedge F_2$,
where we used $\Omega = \frac{1}{3!} f \epsilon_{ijk} \d z^{ijk}$, and $J = i g_{i\bar j} \d z^{i\bar j}$.})
\begin{align}
\d \phi \wedge J \wedge J  & = 3 \left( W_5 + \bar W_5 \right)	\wedge J \wedge J 		\\
& = 	- \frac{3}{4} e^{\phi} \left( F^{0,2}\wedge\Omega + F^{2,0}\wedge\bar\Omega  \right)		\\
& = 	- \frac{3}{2} e^{\phi} F_2\wedge\Omega_R 		  \,.
\end{align}

\bibliography{refs}

\end{document}